\documentclass{ws-procs9x6}

\newcommand{\ffat}[1]{\mbox {\boldmath $#1$}}

\begin{document}

\title{Nuclear Forces and Few-Nucleon Studies Based on Chiral Perturbation
Theory}

\author{W.Gl\"ockle$^1$, E.Epelbaum$^2$, U.G.Mei{\ss}ner$^3$,A.Nogga$^4$,H.Kamada$^5$, H.Wita{\L}a$^6$}

\address{$^1$Inst. Theor. Phys., Ruhr-Universit\"at Bochum, D-44780 Bochum, Germany}
\address{$^2$Jefferson Laboratory, Newport News, VA 23606, USA}
\address{$^3$Universit\"at Bonn, Helmholtz--Institut f\"ur Strahlen-- und Kernphysik (Theory) \\
D-53115 Bonn and FZ J\"ulich, IKP (Theorie), D-52425 J\"ulich, Germany}
\address{$^4$Inst. for Nuclear Theory, University of Washington, Seattle,
  WA 98195, USA}
\address{$^5$Department of Physics, Faculty of Engineering, Kyushu Institute of Technology, \\
1-1 Sensuicho, Tobata, Kitakyushu 804-8550, Japan}
\address{$^6$Jagellonian University, Institute of Physics, Reymonta 4, 30-059 Cracow, Poland}


\maketitle

\abstracts{ After a brief review on the status of few--nucleon
 studies based on conventional nuclear forces, we sketch the concepts of
the effective field theory approach constrained by chiral symmetry
 and its application to nuclear forces. Then first results
 for few--nucleon observables are discussed.}

\section{Introduction}

The basic questions posed in few--nucleon physics have a long tradition.
Already E.Wigner \cite{wigner33} and J.Schwinger \cite{gerjuoy42} 
asked, whether one
can understand the binding energies of the helium nuclei on the basis of
two--nucleon (NN) forces and the Schr\"odinger equation. Over the years,
robust mathematical formulations, which are strictly equivalent to the
Schr\"odinger equation, have been developed. These are the
Faddeev-Yakubovsky equations \cite{faddeev61,yakubovsky67}, the Greens Function Monte Carlo \cite{carlson87},
the stochastic variational \cite{varga95}, the Gaussian basis \cite{kamimura88} 
and
the hyperspherical harmonics \cite{ripelle83} 
methods, which together  with modern NN
forces like AV18 \cite{wiringa95}, Nijm I,II \cite{stoks94} and 
CD-Bonn \cite{machleidt96} allow for an unambiguous
answer: the available NN forces alone underbind light
nuclei. Some examples for theoretical binding energies in comparison to
experimental ones are displayed in Table \ref{table1}. 
 Also in 3N scattering quite a few discrepancies appear  using NN forces
only \cite{glockle96,sekiguchiconf}. A missing dynamical ingredient, naturally
suggested in meson theory, is the three--nucleon (3N) force. First trial
models built around the old Fujita-Miyazawa force \cite{fujita57}, 
a $2\pi$--exchanges
force with an intermediate $\Delta$, are the Tucson-Melbourne (TM)
 \cite{coon01} and the Urbana IX (URB) \cite{pudliner97} forces, 
which contribute additional binding and can be adjusted to the $^3$H
binding energy in conjunction with the available NN forces. 
Then one can
predict binding energies  beyond A=3 and 3N scattering observables.
As an example, we display in Table \ref{table2} the $\alpha$--particle binding energies
for various force combinations, which come rather close to the
experimental value \cite{nogga02b} . For the partially promising results in 3N
scattering we refer the reader to \cite{witala01a,kuros02b} and \cite{sekiguchiconf}. For nuclei beyond
A=4 see \cite{carlson98} and \cite{pieper01a}, where 
extensions to the Urbana~IX model were applied. 
A systematic approach and consistency
between NN and 3N forces are still missing. In the following we sketch the 
concepts of a consistent approach based on effective field theory (EFT) 
constrained  by
chiral symmetry, show resulting nuclear forces and first applications
in the few--nucleon sector. We end with a brief outlook.

\begin{table}[t]
\hspace*{5mm}
\begin{minipage}[t]{50mm}
\tbl{Binding energies in MeV based one  NN forces only \vspace*{1pt}}
{\footnotesize
\begin{tabular}{|l|r|r|r|r|}
\hline
{} &{} &{} &{} &{}\\[-1.5ex]
NNF & $^3$H & $^4$He & $^6$He & $^6$Li  \\[1ex]
\hline
{} &{} &{} &{} &{}\\[-1.5ex]
Nijm & 7.74 & 24.98 & {}    & {}\\[1ex]
CDB  & 8.01 & 26.26 & {}    & {}\\[1ex]
AV18 & 7.60 & 24.10 & 23.90 & 26.90\\[1ex]
\hline
{} &{} &{} &{} &{}\\[-1.5ex]
Exp  & 8.48 & 28.30 & 29.30 & 32.00\\[1ex] 
\hline
\end{tabular}\label{table1} }
\end{minipage}
\hfill
\begin{minipage}[t]{50mm}
\tbl{NNF +3NF predictions for $^4$He\vspace*{1pt}}
{\footnotesize
\begin{tabular}{|l|r|r|}
\hline
{} &{} &{}\\[-1.5ex]
NNF + 3NF & $^3$H & $^4$He\\[1ex]
\hline
{} &{} &{}\\[-1.5ex]
CDB  + TM  & 8.48 & 28.40\\[1ex]
AV18 + TM  & 8.45 & 28.36\\[1ex]
AV18 + URB & 8.48 & 28.50\\[1ex]
\hline
{} &{} &{}\\[-1.5ex]
Exp  & 8.48 & 28.30\\[1ex] 
\hline
\end{tabular}\label{table2} }
\end{minipage}
\end{table}

\section{The concepts}

We start  with a brief reminder. The QCD Lagrangian for massless up and
down quarks is invariant under global flavor $SU(2)_L \times SU(2)_R$ transformations or,
equivalently, under vector and axial vector transformations. This is
called chiral symmetry. Among other facts, the absence of parity
doublets of low mass hadrons suggests that the axial symmetry is
spontaneously broken. The pions are natural candidates for the required
Goldstone bosons. They acquire a nonvanishing mass due to the explicit
symmetry breaking caused by the small up and down quark masses. We are
interested in low energy nuclear physics, where the degrees of freedom are
the composite hadrons. Their interaction has to be described by an
effective Lagrangian, which could not yet be derived from QCD. 
Nevertheless, at least one important requirement is known:
the effective Lagrangian has to be constrained by chiral symmetry and
should include explicitly symmetry breaking parts proportional to powers of
the quark masses. The application we have in mind is for generic
nucleon momenta comparable to the $\pi$--mass and somewhat higher, but
still smaller than the $\rho$--mass. In that case a standard one--boson exchange
picture turns into NN contact forces for the heavy meson exchanges
and only the one-pion exchange is kept explicitly. The
construction of the most general effective Lagrangian out of pion and
nucleon fields constrained by chiral symmetry is nontrivial due to the
fact that no nontrivial linear realization (representation)  of
$SU(2) \times SU(2)$ with pion fields can be formed.
The formalism has been worked out in seminal papers by \cite{weinberg68,coleman69}. The 
chirally invariant expressions are build up out of the nucleon fields and 
covariant derivatives (nonlinear in the pion
fields)  and of the pion and nucleon fields. There is an infinite number of
possible terms, which can be ordered according to the parameter
\begin{equation}
\Delta = d + \frac{1}{2}n -2
\end{equation}
characterizing the vertices. Here $d$ is the number of derivatives and $n$
the number of nucleon field operators. Spontaneously broken chiral symmetry 
enforces $\Delta \ge 0$.

The first few terms for the interacting effective Lagrangian after a $p/m$
expansion (for heavy baryon formalism see \cite{jenkins91}) look like
\begin{eqnarray}
\begin{array}{r}
\mathcal{L}_{eff} \quad = \\
\quad 
\end{array} & & 
\left. \begin{array}{l}
-N^{\dagger} \left[ \frac{g_A}{2F}{\ffat \tau} {\ffat \sigma} \cdot \cdot {\ffat \nabla} {\ffat \pi} + \frac{1}{4F^2}{\ffat \tau} \cdot ({\ffat \pi} \times \dot{{\ffat \pi}}) + \cdots \right] N\\
-\frac{1}{2}C_S\left(N^{\dagger}N\right)\left(N^{\dagger}N\right) - \frac{1}{2}C_T\left(N^{\dagger}{\ffat \sigma}N\right)\left(N^{\dagger}{\ffat \sigma}N\right)
\end{array} \right\} \Delta = 0 \nonumber\\
& & 
\left. \begin{array}{l}
+\frac{1}{F^2}N^{\dagger}\left[-2c_1m_{\pi}^2\pi^2+c_3\partial_{\mu}{\ffat \pi}\partial^{\mu}{\ffat \pi} \right.\\
-\left.\frac{1}{2}c_4\varepsilon_{ijk}\varepsilon_{abc}\sigma_i\tau_a(\nabla_j\pi_b)(\nabla_k\pi_c) + \cdots \right]N\\
+\frac{D_1}{F}(N^{\dagger}N)(N^{\dagger}{\ffat \sigma}\cdot{\ffat \tau}N) \cdot \cdot {\ffat \nabla}{\ffat \pi}\\
+ E_1(N^{\dagger}N)(N^{\dagger}{\ffat \sigma}N)^2 
\end{array} \qquad \quad \right\} \Delta = 1 \nonumber\\
& & 
\left. \begin{array}{l}
-\frac{1}{2}C_1\left[(N^{\dagger}{\ffat \nabla}N)^2 + ({\ffat \nabla}N^{\dagger}N)^2\right]\\
+ \cdots \\
+ C_7(\partial_iN^{\dagger}\sigma_l\partial_iN)(N^{\dagger}\sigma_lN)
\end{array} \qquad \qquad \qquad \quad \right\} \Delta = 2 \nonumber\\
& & + \cdots 
\end{eqnarray}
The terms are grouped according to $\Delta =0,1$ and 2. The parameters
of $\mathcal{L}_{eff}$, 
the so called low--energy constants (LEC's), can be partitioned in several
groups: 
some can be determined
in the $\pi$-N system ($g_A, F,c_1,c_3,c_4$) and others from nucleonic 
systems only ($C_S,C_T,E_1,C_1,... C_7$). The constant $D_1$ also affects the
NN$\pi$ system. All these constants are of course not determined  by
chiral symmetry, but have to be adjusted to experimental  data.

\section{Nuclear Forces}

To arrive at nuclear forces the pion degrees of freedom have to be
eliminated. We use an unitary transformation of the field theoretical
Hamiltonian \cite{epelbaumphd,epelbaoum98b} going back to Okubo \cite{okubo54}. That transformation decouples
the purely nucleonic Fock space from the one, which includes at least one
pion. That elimination process is controlled by the 
low--momentum expansion \cite{epelbaumphd}.
A resulting NN force $V$  receives contributions 
of the order
\begin{equation}
V \sim ( Q/\Lambda)^{\nu}
\end{equation}
where Q is a generic external momentum and $\Lambda$ is the mass scale, which
enters the (renormalized) values of the LEC's.
The power $\nu$ is given by
\begin{equation}
\nu = -4 + 2N + 2L + \sum_i V_i \Delta_i,
\end{equation}
here L is the number of loops and $V_i$ the number of vertices of type $i$.
Therefore, if one sticks to momenta $Q$ such that $Q \ll \Lambda$ and
$Q/\Lambda$ is a small quantity, the
effects of nuclear forces decrease with increasing $\nu$.
Clearly, since $\Delta \ge 0 $, one has $\nu \ge 0$. Thus, if one
wants to derive nuclear forces in the low--$Q$ regime with a given accuracy,
one needs only a finite number of terms in $\mathcal{L}_{eff}$ 
with the smallest $\Delta$'s and the finite number of pions.

\begin{figure}[t]
\centerline{\epsfxsize=60mm\epsfbox{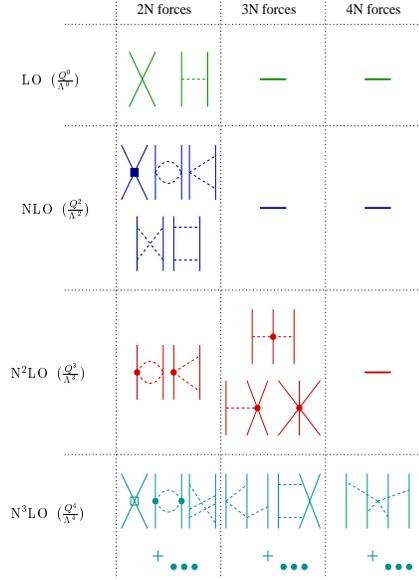}}
\vskip -0.4 true cm
\caption{\label{fig1}Hierarchy of nuclear forces.}
\end{figure}

The first few orders for nuclear forces can easily be listed using  the terms in
Eq.~(2). The leading order (LO) for $\nu=0 $
and $N=2$ requires $L=0$ and $\Delta_i=0$. This leads to two types of vertices
with $d=1$, $n=2$ and $d=0$, $n=4$, which occur in the one-pion exchange and two
contact forces, respectively. $\nu=$1 does not exist and the next-to-leading
order (NLO) with $\nu=2$ and $N=2$ requires either $L=1$, $\Delta_i=0$ or $L=0$, 
$\Delta_i=2$. In the first case one can form various types of 
$2\pi$--exchange processes with $d=1$, $n=2$ vertices; in the second case one
encounters additional NN contact forces with 7 different types of vertices 
with $d=2$, $n=4$. At NNLO ($\nu=3$) additional 
$2\pi$--exchange NN forces with higher order vertices occur. Also the first
nonvanishing 3N forces of three different topologies show up. They are 
without loops ($L=0$) and with $\Delta \le 1$ vertices. A
$2\pi$--exchange process, a $1\pi$--exchange between a NN contact force and the
third nucleon and a pure 3N contact force are of this form. 
Each of the latter two
depends on one unknown LEC, whereas the $2\pi$--exchange is
parameter--free in the sense that the LEC's are determined in the 
$\pi N$--system. That hierarchy of nuclear forces is illustrated in Fig. \ref{fig1}.
Thus, chiral EFT provides a natural explanation why
3N forces are less important than 2N forces, 4N forces are less
important than 3N forces, etc. We would also like to emphasize that all
forces are analytically given and we refer the reader to \cite{epelbaum02a}.

The derived nuclear forces are valid only at low $Q$ and have
unphysical behavior at large $Q$ (they grow with increasing momenta).
The Schr\"odinger equation needs therefore to be regularized.
This is achieved by introducing $V_{reg}$ according to 
\begin{equation}
V_{reg}({\bf p}',{\bf p})= e^{-({p'}^2/\Lambda^2)^2} V({\bf p}',{\bf p})e^{-(p^2/\Lambda^2)^2}\,.
\end{equation}
$\Lambda$ should be not too small in order not to cut off the physics of
the $\pi$-exchanges and not too large in order to exclude uncontrolled high--energy physics.
It turned out that
\begin{equation}
500 \mbox{ MeV/c} \le \Lambda \le 600 \mbox{ MeV/c}
\end{equation}
is a good choice. The $\Lambda $-dependence is expected to get weaker
with increasing order in the expansion \cite{lepage97,gegelia01}. Notice that the
original idea and the first applications of chiral perturbation theory to nuclear systems go 
back to \cite{weinberg91,ordonez96,kaiser97}. 
The higher orders have been worked out by N.~Kaiser \cite{kaiser02} and 
also applied in \cite{entem03a}.

\section{Application in the Few-Nucleon Sector}

\begin{figure}[t]
\centerline{\epsfxsize=80mm\epsfbox{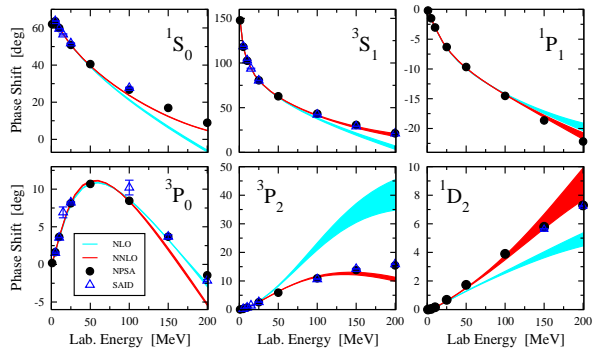}}
\vskip -0.4 true cm
\caption{\label{fig2}NN phase shifts for NLO versus NNLO.}
\end{figure}

\begin{figure}[t]
\centerline{\epsfxsize=80mm\epsfbox{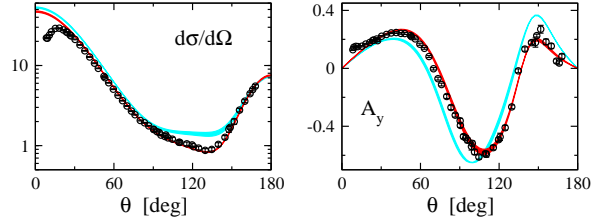}}
\vskip -0.4 true cm
\caption{\label{fig3}pd elastic observables at 65 MeV. Curves as in Fig.2. For data see \protect \cite{epelbaum02c}}
\end{figure}

\begin{figure}[t]
\centerline{\epsfxsize=80mm\epsfbox{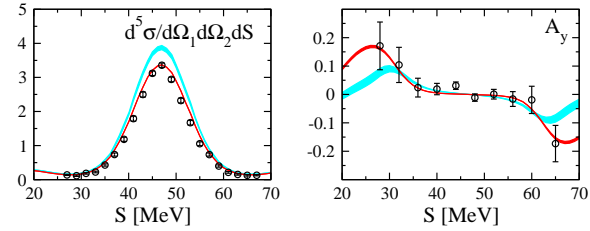}}
\vskip -0.4 true cm
\caption{\label{fig4} pd break up data versus theory. Curves as in Fig.~2. For data see \protect \cite{epelbaum02c}}
\end{figure}

\begin{figure}[t]
\hspace*{10mm}
\begin{minipage}[t]{40mm}
\centerline{\epsfxsize=36mm\epsfbox{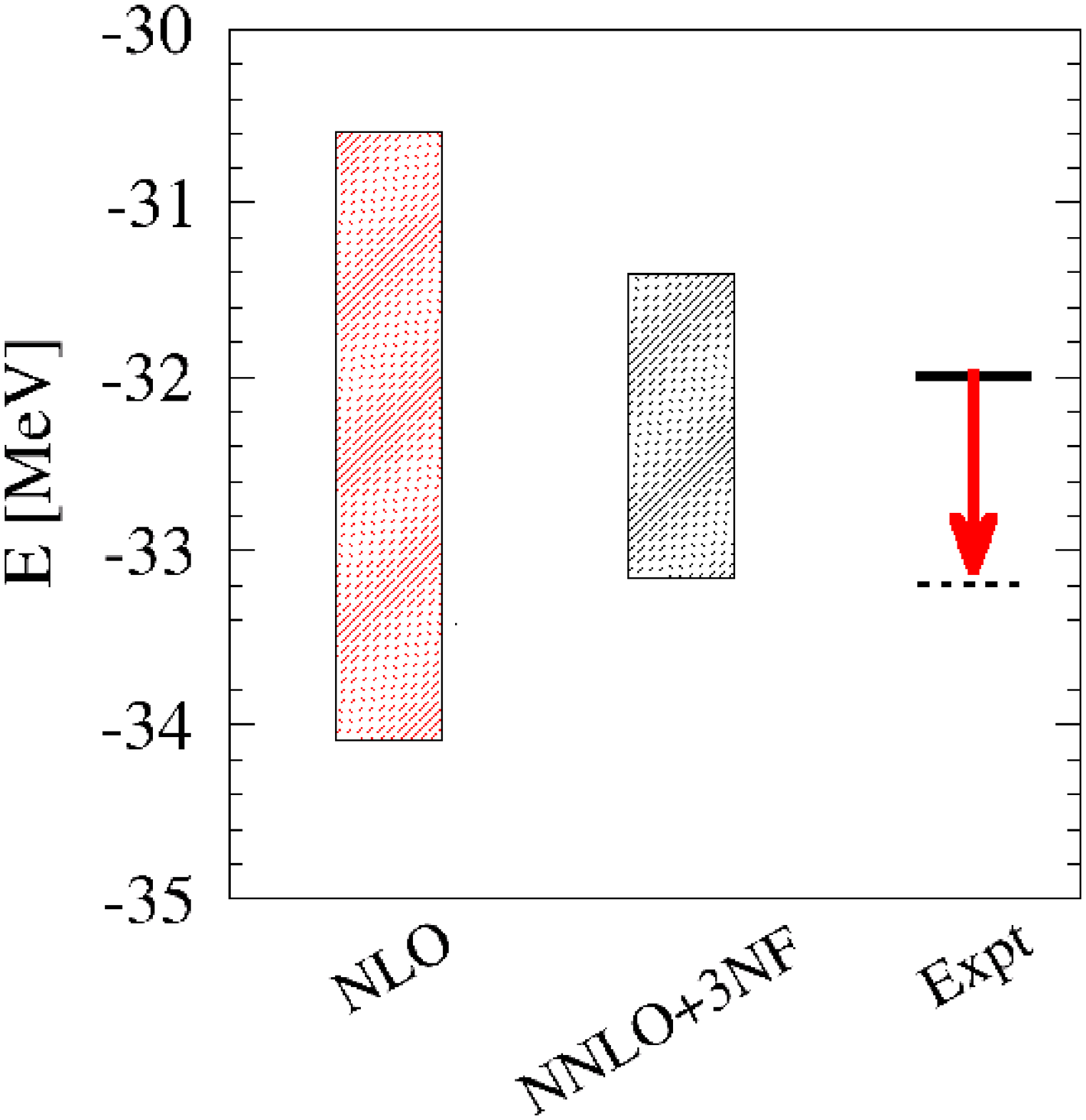}}
\end{minipage}
\hspace*{10mm}
\begin{minipage}[t]{40mm}
\centerline{\epsfxsize=40mm\epsfbox{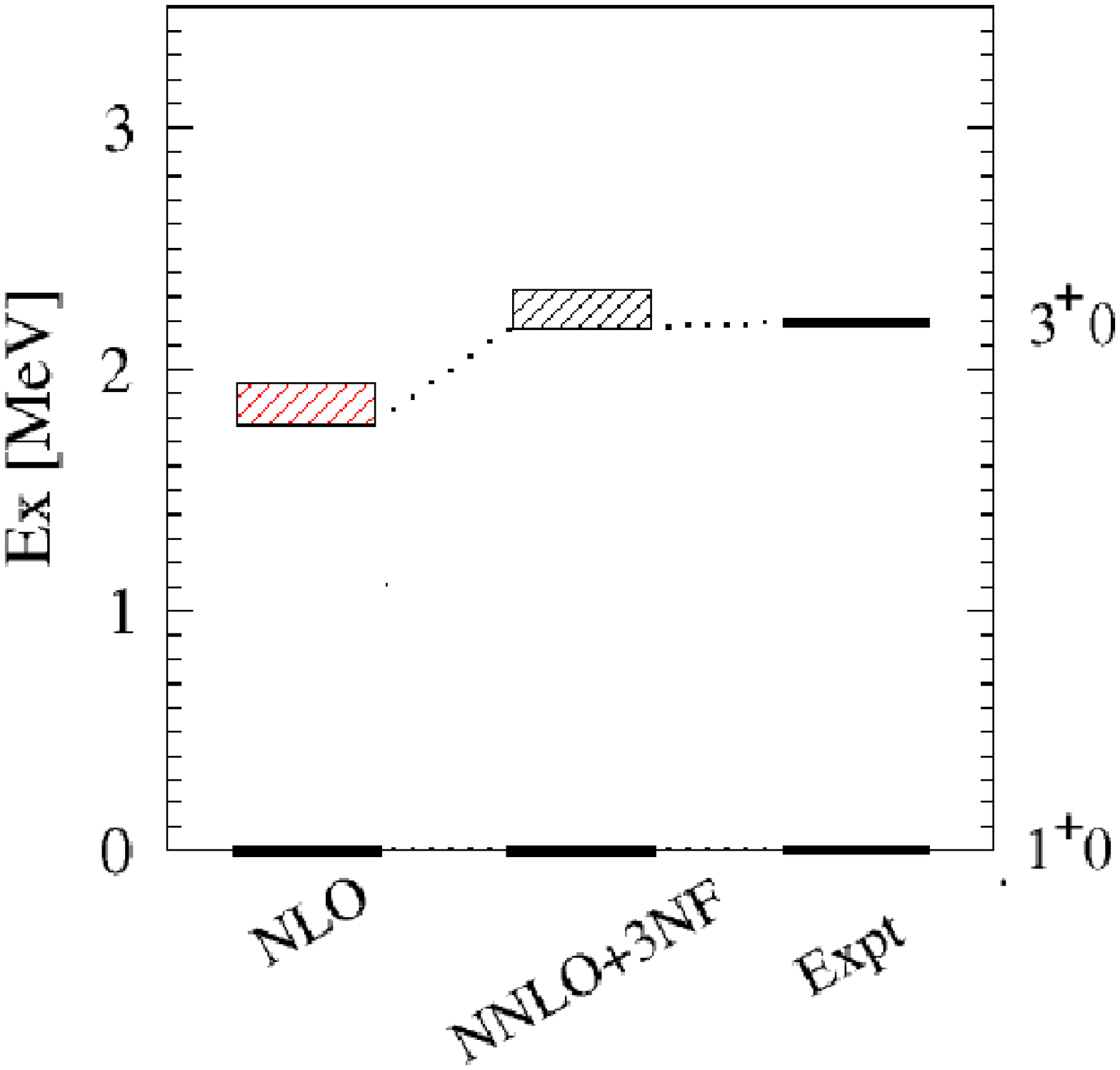}}
\end{minipage}
\vskip -0.4 true cm
\caption{\label{fig5} Predictions for the $^6$Li ground and excited state.}
\end{figure}

The first step is the adjustment of the LEC's.
There are 2 (9) such constants  at LO (NLO), which are adjusted to the NN
S-- and P--wave phase shifts. The only additional LEC's appearing at
NNLO are $c_{1,3,4}$, which can in principle be taken from the 
pion--nucleon system (see  \cite{epelbaum02a} for more details). 
Our results for selected partial waves are 
illustrated in Fig.~\ref{fig2}. 
In the 3N system we adjust the two LEC's
entering the 3N force to the $^3$H binding energy and
the doublet nd scattering length $a_{nd}$ \cite{epelbaum02c}. Then, up to and including NNLO, all
constants are fixed. We now display some results and refer the
reader to \cite{epelbaum02c} for more details.
The $\alpha$-particle binding energy for $\Lambda$ = 500 (600) MeV/c turns
out to be 29.51 (29.98) MeV which comes close to the "experimental" value
29.8 MeV (This is a corrected value for np forces only).
Also the results for 3N scattering look mostly promising and we display
 a few examples in Figs.~\ref{fig3} and \ref{fig4}. Finally we show in Fig.~\ref{fig5} 
an application to $^6$Li which
has  been elaborated in the no--core shell model  framework \cite{nogga03b}.

\section{Summary and Outlook}

 EFT and chiral symmetry is a systematic path towards
nuclear forces. The nuclear forces are built out of multi-pion
exchanges, which are parameter  free and a string of contact
forces, which parameterize the not yet understood short range
physics. 3N forces are consistent to 2N forces.
The dominance of 2N forces over 3N forces is
naturally explained in this theoretical framework.
The regularization of the pion loops, which removes uncontrolled high
energy (short range) physics, has recently been formulated using the spectral
function representation \cite{epelbaum03a,epelbaum03b} and the application to
NN forces at NNNLO($\nu$=4) is in progress
\cite{epelbaumprep}. At this order quite a few new 3N forces 
appear. They are parameter free and, therefore, their effects on 
few--nucleon observables will be of special interest. 
Relativistic  corrections can be taken into account and are 
expected to converge rapidly since the nucleon momenta stay well below the nucleon
mass.


\begin{thebibliography}{10}
\expandafter\ifx\csname url\endcsname\relax
  \def\url#1{\texttt{#1}}\fi
\expandafter\ifx\csname urlprefix\endcsname\relax\def\urlprefix{URL }\fi

\bibitem{wigner33}
E.~Wigner, Phys. Rev. 43 (1933) 252.

\bibitem{gerjuoy42}
E.~Gerjuoy, J.~Schwinger, Phys. Rev. 61 (1942) 138.

\bibitem{faddeev61}
L.~D. Faddeev, Sov. Phys. JETP 12 (1961) 1014.

\bibitem{yakubovsky67}
O.~A. Yakubovsky, Sov. J. Nucl. Phys. 5 (1967) 937.

\bibitem{carlson87}
J.~Carlson, Phys. Rev. C 36 (1987) 2026.

\bibitem{varga95}
{Y. Suzuki}, {K. Varga}, Phys. Rev. C 52 (1995) 2885.

\bibitem{kamimura88}
M.~Kamimura, Phys. Rev. A 38 (1988) 621.

\bibitem{ripelle83}
M.~Fabre~de~la Ripelle, Ann. Phys. (N.Y.) 147 (1983) 281.

\bibitem{wiringa95}
{R.~B. Wiringa}, {V.~G.~J. Stoks}, {R. Schiavilla}, Phys. Rev. C 51 (1995) 38.

\bibitem{stoks94} 
{V.~G.~J. Stoks} {\it et. al.},
  Phys. Rev. C 49 (1994) 2950.

\bibitem{machleidt96}
{R. Machleidt}, {F.~Sammarruca}, {Y.~Song}, Phys. Rev. C 53 (1996) R1483.

\bibitem{glockle96}
W.~Gl\"ockle {\it et. al.}, Phys. Rep. 274
  (1996) 107.

\bibitem{sekiguchiconf}
K.~Sekiguchi, contribution to this conference.

\bibitem{fujita57}
{J. Fujita and H. Miyazawa}, Progress of Theor. Phys. 17 (1957) 360.

\bibitem{coon01}
{S.~A. Coon}, {H.~K. Han}, Few Body Systems 30 (2001) 131.

\bibitem{pudliner97}
{B.~S. Pudliner} {\it et. al.}, Phys. Rev. C 56 (1997) 1720.

\bibitem{nogga02b}
A.~Nogga {\it et. al.}, Phys. Rev. C 65 (2002)
  054003.

\bibitem{witala01a}
H.~Wita{\l}a {\it et. al.}, Phys. Rev. C 63 (2001) 024007.

\bibitem{kuros02b}
{{J. Kuro\'{s}-\.{Z}o{\l}nierczuk} {\it et. al.}, Phys. Rev. C 66 (2002) 024004.

\bibitem{carlson98}
J.~Carlson, R.~Schiavilla, Rev. of Mod. Phys. 70~(3) (1998) 743.

\bibitem{pieper01a}
S.~C. Pieper {\it et. al.}, Phys. Rev. C 64
  (2001) 014001.

\bibitem{weinberg68}
S.~Weinberg, Phys. Rev. 166 (1968) 1568.

\bibitem{coleman69}
S.~Coleman, J.~Wess, B.~Zumino, Phys. Rev. 177 (1969) 2239.

\bibitem{jenkins91}
E.~Jenkins, A.~V. Manohar, Phys. Lett. B 255 (1991) 558.

\bibitem{epelbaumphd}
E.~Epelbaum, Ph.D. thesis, Ruhr-Universit\"at, Bochum (2000).

\bibitem{epelbaoum98b}
E.~Epelbaoum, W.~Gl\"ockle, U.-G. Mei{\ss}ner, Nucl. Phys. A637 (1998) 107.

\bibitem{okubo54}
{S.~Okubo}, Progr. Theor. Phys. 12 (1954) 603.

\bibitem{epelbaum02a}
E.~Epelbaum {\it et. al.}, Eur. Phys. J. A 15 (2002) 543.

\bibitem{lepage97}
G.~P. Lepage, nucl-th/9706029.

\bibitem{gegelia01}
J.~Gegelia, G.~Japaridze, Phys. Lett. B 517 (2001) 476.

\bibitem{weinberg91}
S.~Weinberg, Nucl. Phys. B363 (1991) 3.

\bibitem{ordonez96}
C.~Ord\'{o}\~{n}ez, L.~Ray, U.~van Kolck, Phys. Rev. C 53 (1996) 2086.

\bibitem{kaiser97}
N.~Kaiser, R.~Brockmann, W.~Weise, {Nucl. Phys.} A625 (1997) 758.

\bibitem{kaiser02}
N.~Kaiser, {Phys. Rev. C} 61 (1999) 014003; {Phys. Rev. C} 62 (2000) 024001; 
{Phys. Rev. C} 64 (2001) 057001.

\bibitem{entem03a}
D.~R. Entem, R.~Machleidt, Phys. Rev. C 68 (2003) 041001(R).

\bibitem{epelbaum02c}
E.~Epelbaum {\it et. al.}, Phys. Rev. C 66 (2002) 064001.

\bibitem{nogga03b}
A. Nogga {\it et. al.}}, to appear in the proceedings of the 17th
  International Conference on Few-Body Problems in Physics, Durham, NC, 2003.

\bibitem{epelbaum03a}
E.~Epelbaum, W.~Gl\"ockle, U.-G. Mei{\ss}ner, nucl-th/0304037.

\bibitem{epelbaum03b}
E.~Epelbaum, W.~Gl\"ockle, U.-G. Mei{\ss}ner, nucl-th/0308010.

\bibitem{epelbaumprep}
E.~Epelbaum {\it et. al.}, in preparation.

\end{thebibliography}
\end{document}